# Experimental Determination of a Single Atom Ground State Orbital through Hyperfine Anisotropy

*Laëtitia Farinacci[†1], Lukas M. Veldman[†1], Philip Willke[2], Sander Otte[1,*]*

1 Department of Quantum Nanoscience, Kavli Institute of Nanoscience, Delft University of Technology, 2628 CJ Delft, the Netherlands.
2 Physikalisches Institut, Karlsruhe Institute of Technology, Karlsruhe, Germany
*E-mail: corresponding author: a.f.otte@tudelft.nl
[†] These authors contributed equally

**Historically, electron spin resonance (ESR) has provided excellent insight into the electronic, magnetic, and chemical structure of samples hosting spin centers. In particular, the hyperfine interaction between the electron and the nuclear spins yields valuable structural information of these centers. In recent years, the combination of ESR and scanning tunneling microscopy (ESR-STM) has allowed to acquire such information on individual spin centers of magnetic atoms bound atop a surface, while additionally providing spatial information about the binding site. Here, we conduct a full angle-dependent investigation of the hyperfine splitting for individual titanium atoms on MgO/Ag(001) by measurements in a vector magnetic field. We observe strong anisotropy in both the g-factor and the hyperfine tensor. Combining the results of the hyperfine splitting with the symmetry properties of the binding site obtained from STM images and a basic point charge model allows us to predict the shape of the electronic ground state configuration of the titanium atom. Relying on experimental values only, this method paves the way for a new protocol for electronic structure analysis for spin centers on surfaces.**

KEYWORDS: Scanning Tunneling Microscopy, Electron Spin Resonance, Hyperfine Interaction, Vector magnetic field, Single-Atom Magnetism, Magnetic Sensing

For decades, nuclear spins have constituted an excellent resource to gain information on the atomic scale [1]. In recent years, advances in many different architectures, including nitrogen-vacancy-centers in diamond [2], molecular break junctions [3] and phosphorous donors in silicon [4], even allowed to address them on an individual level. This effort is mainly driven by their prospect as a future building block in quantum information processing and sensing [5]. However, nuclear spins have been used for even longer to gain structural and electronic information about materials in bulk experiments. The nuclei can be probed directly using nuclear magnetic resonance measurements as well as indirectly via ESR since the magnitude and anisotropy of the hyperfine interaction is reflected in properties of the electron cloud surrounding the nucleus [1].

The combination of electron spin resonance and scanning tunneling microscopy (ESR-STM) has opened a novel platform to access single nuclear spins of atoms on surfaces [6-9]. Most strikingly, both spatial and magnetic information can be obtained by the two techniques simultaneously, providing unique access to hyperfine interaction on the atomic scale. Previous experiments showed that the hyperfine interaction of individual titanium (Ti) atoms on a bilayer of magnesium oxide (MgO) strongly depends on the binding side [7]. Initial experiments hinted towards a strong anisotropic hyperfine interaction on all binding sides. However, these measurements were performed in one magnetic field direction only; this limited the electronic structure analysis and required the additional help of density functional theory (DFT) to interpret the data [7].

Here, we perform ESR-STM measurements of individual Ti atoms on a bridge binding side of MgO in a vector magnetic field. We demonstrate that the hyperfine tensor has distinctly different values along its principal axes than reported previously [7]. Combining the results from the hyperfine analysis with properties of the symmetry group of the atom's binding site derived from STM and a basic point charge model allows us to predict the shape of the ground state orbital of the atom without the use of first-principle calculations such as DFT.

Experiments were conducted in a commercial STM system (Unisoku USM1300) equipped with a vector-magnetic field (Fig. 1a) and at a temperature of 1.5 K. The measurements were performed on well-isolated individual Ti atoms adsorbed on two atomic layers of MgO grown on a Ag(100) substrate. These titanium atoms were found to be hydrogenated by residual hydrogen in the vacuum chamber [10] effectively reducing them to $Ti^{3+}$ with spin $S = 1/2$. Figure 1b shows a STM topography of a single Ti atom. For ESR experiments, a radiofrequency (RF) voltage $V_{RF}$ is applied to the STM tip in addition to the DC bias voltage $V_{DC}$. This RF voltage can drive transitions between the two lowest lying spin states of the Ti atom, which is subsequently detected by changes in the tunnel current $\Delta I$ via magneto-resistive tunneling. For the latter, a magnetic STM tip is employed that is created by transferring several Fe atoms from the surface to the STM apex. We study Ti atoms adsorbed on O-O bridge sites, which come in two equivalent orientations as shown in Figure 1c: 'horizontal' and 'vertical', which have an in-plane magnetic field angle with respect to the crystal lattice of 14° and 76°, respectively. This leads effectively to two different orientations of the in-plane field and thus allows for a 3-dimensional mapping of the hyperfine interaction by rotating the magnet only in a single plane (see supplementary S1).

In accordance with Ref. [7], we can identify three different configurations of the Ti nuclear spin. In Figure 2, we display different ESR spectra measured above atoms adsorbed on vertical bridge sites; we observe a single ESR resonance for $^{46}$Ti, $^{48}$Ti and $^{50}$Ti ($I = 0$), six resonances for $^{47}$Ti ($I = 5/2$), and eight for $^{49}$Ti ($I = 7/2$). In line with previous experiments, we observe a variation of the overall signal intensity for different magnetic field angles [11]. Interestingly, for the isotopes carrying a non-zero nuclear spin, the different peaks are well-resolved when the external field is along the sample plane, with a splitting around ~65 MHz, while they seem to merge when the field is aligned in the out-of-plane direction, with a ~20 MHz splitting. This strong anisotropy of the hyperfine splitting is remarkable and could not be accurately determined with measurements performed along a single field direction [7].

In Fig. 3, we map the full evolution of the ESR spectra as a function of $\theta$, the angle of the magnetic field with respect to the surface normal, for two perpendicular rotation planes. Panel (a) shows data taken on a $^{49}$Ti atom on a vertical bridge site, meaning that the in-plane field makes a 14° angle with the x-axis. The data exhibits strong anisotropic behaviour, with almost complete suppression of the hyperfine splitting for the out-of-plane field direction. All data in this panel was acquired with the same microtip, and, by measuring for each data point a reference spectrum on a $^{48}$Ti atom, we can ensure that the influence of the tip field is negligible (see supplementary S1).

We performed the same experiment on another $^{49}$Ti atom adsorbed on a horizontal bridge site, with a different microtip but that is again kept the same for the whole data set (see Figure 3b). Also here, we observe anisotropic behavior of the hyperfine splitting, though much less dramatic than for the vertical binding site. The evolution of the hyperfine splitting can be quantified by fitting each spectrum with several Fano functions (see supplementary S1) and is shown in Figure 3c for both adsorption sites. The evolution of the hyperfine splitting is continuous and mirror-symmetric, indicating that the sign of the magnetic field along any direction is irrelevant. We note that the observed symmetry axis is rotated by ~10° with respect to the magnet axes. We discuss possible origins for this rotation in supplementary S2. From the anisotropic evolution of the hyperfine splitting in Figure 3c we can already infer that the extent of the ground state orbital, which scales the hyperfine splitting via the magnetic dipole-dipole interaction, is likely to be similar in two directions (out-of-plane and one in-plane) and differs substantially in the other (in-plane) one.

The anisotropy of the hyperfine splitting is closely related to that of the g-factor. The latter had already been observed for Ti on MgO/Ag(100) [11-13]. The hyperfine interaction entails three different interactions: a dipole-dipole interaction between the electron and nuclear spins, a Fermi contact interaction that scales with the electron density at the position of the nucleus, and an orbit dipolar interaction that couples the nuclear spin and angular momentum of the unpaired electron. Spin-orbit coupling leads to a partially unquenched angular momentum which couples to the electron spin. Treating this effect up to second order with perturbation theory, one can write a spin Hamiltonian in which, in all generality, *g* and *A* are tensors [1]:

$$\widehat{H}_{spin} = \mu_B \boldsymbol{B} \cdot \boldsymbol{g} \cdot \widehat{\boldsymbol{S}} + \widehat{\boldsymbol{S}} \cdot \boldsymbol{A} \cdot \widehat{\boldsymbol{I}}. \tag{1}$$

The symmetry of the adsorption site often lowers the degree of anisotropy of these tensors for a particular set of axes $(x, y, z)$. In fact, in traditional ESR spectroscopy, analysis of the hyperfine anisotropy in a vector magnetic field is used to determine the symmetry of the crystal field around the investigated species [1, 14, 15]. This powerful method compensates for the lack of spatial resolution in these ensemble measurements and permits to even observe effects due to hybridization with ligand orbitals [16]. In our case, the combination of ESR with STM allows us to measure ESR spectra of single atoms, while the symmetry of the adsorption site can be exactly determined by STM. As we show, we can thus perform an all-experimental electronic analysis in order to determine the shape of the ground state orbital, a quantity that has been long elusive for experimentalists.

The adsorption site of the atom has a $C_{2v}$ symmetry (see Figure 4) so that $\boldsymbol{g}$ and $\boldsymbol{A}$ are vectors along the principal axes $(x, y, z)$ of the crystal lattice [16]. In the presence of an external magnetic field that has $(l, m, n)$ directional cosines with respect to these axes, the effective $g$ and $A$ parameters are given by [1]:

$$g = \sqrt{(lg_x)^2 + (mg_y)^2 + (ng_z)^2}, \tag{2}$$

$$A = \frac{1}{g}\sqrt{(lg_x A_x)^2 + (mg_y A_y)^2 + (ng_z A_z)^2}. \tag{3}$$

Using these two equations we first determine the effective g values for the vertical and horizontal bridge sites corresponding to different in-plane fields. We find that the vector $\boldsymbol{g}$ is completely anisotropic with $g_x = 1.702 \pm 0.004$, $g_y = 1.894 \pm 0.004$ and $g_z = 2.011 \pm 0.015$. These values are in good agreement with the literature values [11] and the small deviations can be explained by the presence of a small residual tip field. Since this tip field has been carefully accounted for by J. Kim et al., we use in the following their reported g-values [11]. Next, we fit the data of Figure 3c to obtain the values of the hyperfine splitting, first along our field directions and, finally, along the lattice directions (see supplementary S2). We here find $A_x = (68 \pm 4)$ MHz, $A_y = (18 \pm 4)$ MHz, and $A_z = (19 \pm 4)$ MHz. The minima of the two data sets are each a measure of $A_z$, however, they are not exactly equal. We attribute the difference, which has been taken into account for the estimation of the error in $A_z$, to small variations in the local electric field surrounding the two atoms. Statistical variations of the g factor of Ti atoms adsorbed on oxygen sites were indeed also observed and attributed to the same origin [13]. The errors for the in-plane components are dominated by the uncertainty concerning the tilt of the in-plane field with respect to the crystal lattice (see supplementary S2).

Once both the values of $\boldsymbol{g}$ and $\boldsymbol{A}$ are determined, we can investigate how these relate to the $d^1$ ground state configuration of the $Ti^{3+}$. The corresponding energy diagram for $C_{2v}$ symmetry is displayed in Figure 4b [16]. The order of the excited states is arbitrarily chosen and bears no influence on the analysis. The ground state orbital is a superposition of $d_{x^2-y^2}$, $d_{z^2}$ and $4s$ orbitals and our study revolves around determining the values of their respective weights $c_1$, $c_2$ and $c_s$, which satisfy the normalization equation $c_1^2 + c_2^2 + c_s^2 = 1$. The molecular coefficients $\alpha$, $\beta$, $\gamma_1$, $\gamma_2$ and $\delta$ quantify the hybridization of the $d$ levels with ligand orbitals, which we assume to be small – these coefficients are therefore expected to be close to 1.

In C$_{2v}$ symmetry, the electronic configuration of the $d$ levels causes anisotropy of $\boldsymbol{g}$ in the following way [16]:

$$\Delta g_x = g_x - g_0 = -2\alpha^2(c_1 + \sqrt{3}c_2)^2 K_2 \tag{4}$$

$$\Delta g_y = g_y - g_0 = -2\alpha^2(c_1 - \sqrt{3}c_2)^2 K_3 \tag{5}$$

$$\Delta g_z = g_z - g_0 = -8\alpha^2 c_1^2 K_1 \tag{6}$$

Where $g_0 = 2.0023$, $K_1 = \beta^2 \xi / \Delta(a_2)$, $K_2 = \gamma_2^2 \xi / \Delta(b_2)$, and $K_3 = \gamma_1^2 \xi / \Delta(b_1)$, with $\xi$ being the spin-orbit coupling constant and $\Delta(a_2)$ [$\Delta(b_2)$, $\Delta(b_1)$] the energy difference between the excited state $a_2$ [$b_2$, $b_1$] and ground state $a_1$ (see Figure 4b). As for the A vector we have:

$$\Delta A_i = A_i - A_{mean} = P\alpha^2 f_i(c_1, c_2, K_1, K_2, K_3) \tag{7}$$

where $i = x, y, z$, $A_{mean} = \frac{1}{3}(A_x + A_y + A_z)$, $P = g_0 g_N \mu_N \mu_B \langle r^{-3} \rangle$ ($g_N$: nuclear g factor, $\mu_B$: electron Bohr magneton, $\mu_N$: nuclear Bohr magneton) scales with the radial extent of the electronic wave-function via $\langle r^{-3} \rangle$, and $f_i$ are functions whose full expressions can be found in the supplementary S3. These equations, along with the normalization condition for $c_1, c_2$ and $c_s$ above, allow us to calculate the anisotropy of $\boldsymbol{g}$ and $\boldsymbol{A}$ for a given set of parameters ($P, \alpha, c_1, c_2, c_s$) and therefore identify all sets of parameters that could, from a symmetry argument, describe our system. We find that more than one set of parameters can lead to the experimentally observed $\boldsymbol{g}$ and $\boldsymbol{A}$ (see supplementary S3). Consequently, we employ a basic point charge model (supplementary S4) that allows us to discriminate the different solutions by their Coulomb interaction. The lateral positions of the atoms are determined experimentally by atomic resolution STM images. The positions in the $z$-direction of the Ti and H atoms are estimated, but we ensure the robustness of the model against variations of these parameters. The state with the lowest Coulomb energy is shown in Fig. 4c. It consists of a superposition of the $d_{x^2-y^2}$ (74%) and $d_{z^2}$ (26%) orbitals in very good agreement with results obtained from DFT calculations [7]. This is quite remarkable, since our electronic structure analysis is solely based on experimental data assisted by the symmetry group of the surface and a basic point charge model. However, our model cannot discriminate between different values of $c_s$ which scales the admixture of the $4s$ orbital (see supplementary S3). Nevertheless, we show that additional admixture of $c_s$ merely influences the shape of the orbital by reducing the size of the central ring that points toward the neighboring O atoms (see supplementary S4).

In summary, this work illustrates how an analysis of the anisotropic hyperfine interaction can be exploited in order to gain an in-depth knowledge about the shape of the ground state orbital. Crucial for this method is the addition of binding site information derived from STM, that we process in a basic point charge model. Since this protocol can be applied to other spin systems on surfaces in a straightforward manner, it paves the way to determine the spin ground states of atoms and molecules on surfaces and constitutes an independent method that more elaborate theoretical methods such as DFT can be benchmarked against.

While writing this manuscript, we became aware of a similar experiment performed in another group [17]. Overall, their results agree very well with those presented here.

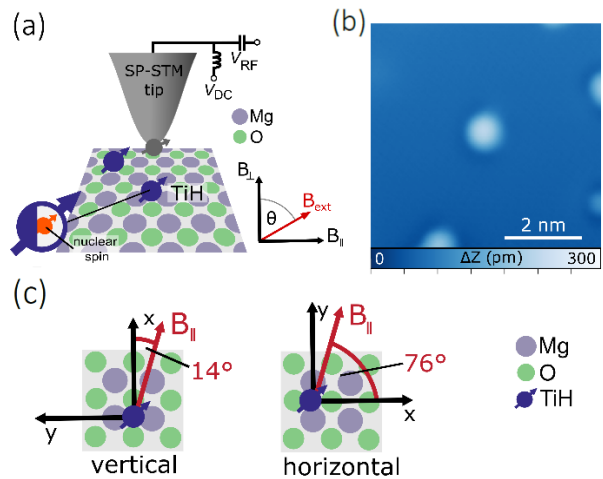

**Figure 1**. Electron spin resonance in a scanning tunneling microscope with a vector magnet. (a) Schematic of the experiment. (b) Topography image of a TiH atom on MgO ($I = 20$ pA, $V_{DC} = 60$ mV). (c) We study TiH atoms adsorbed on two equivalent bridge sites, vertical and horizontal, which effectively correspond to two different directions of the in-plane field $B_\parallel$.

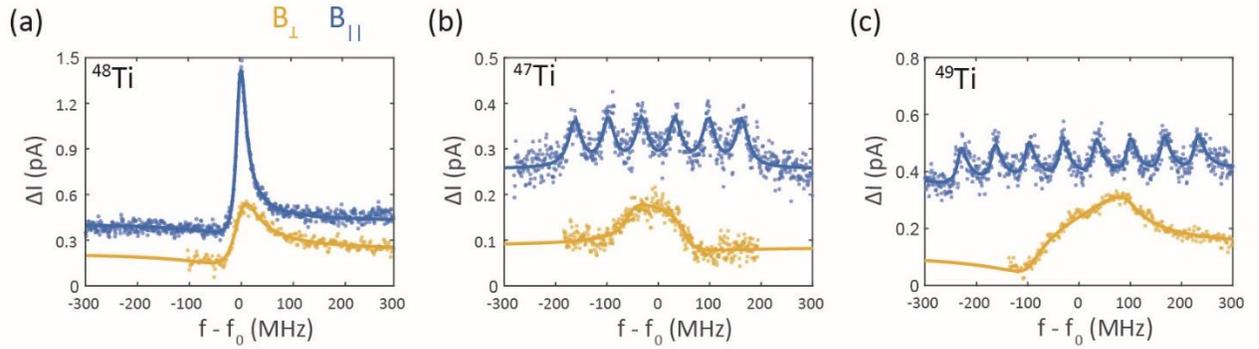

**Figure 2.** ESR spectra of different Ti isotopes (a, b, c) adsorbed on vertical bridge sites in an external magnetic field pointing in-plane (blue) and out-of-plane (orange). Traces were offset with respect to each other for clarity. Experimental parameters: $V_{DC} = 60$ mV, $I = 8$ - $10$ pA, $V_{RF} = 45$ - $57$ mV, $|B_{ext}| = 0.86$ - $1.037$ T, $f_0 = 24.10$ - $24.48$ GHz.

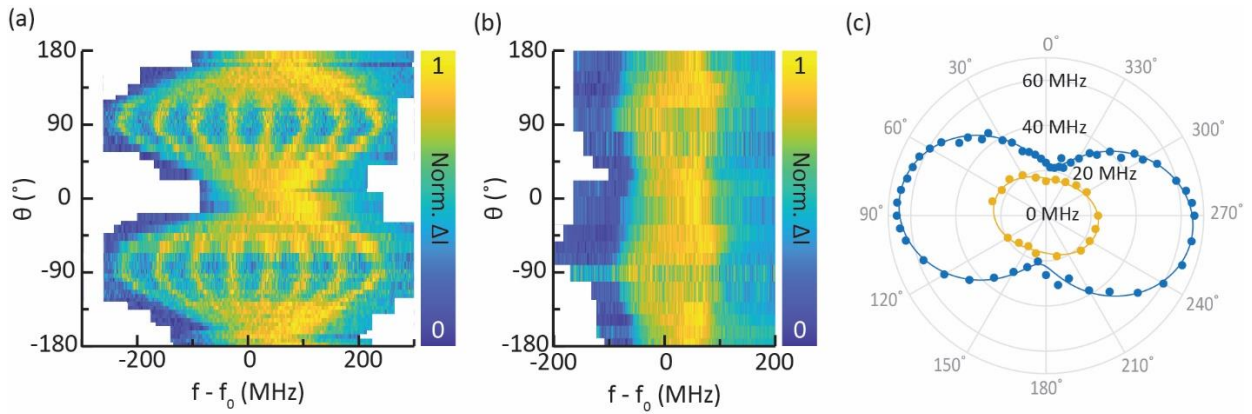

**Figure 3.** Hyperfine splitting in a vector magnetic field of $^{49}$Ti adsorbed on a vertical (a) and horizontal bridge site (b). The data in (c) is obtained by fitting each spectrum in (a) (blue dots) and (b) (yellow dots) with a sum of Fano functions (see supplementary S1), the error bars corresponding to the standard deviation of the fits are smaller than the markers' size. Fits to the experimental data (blue and yellow lines) are based on equations (2) and (3) (see supplementary S2).

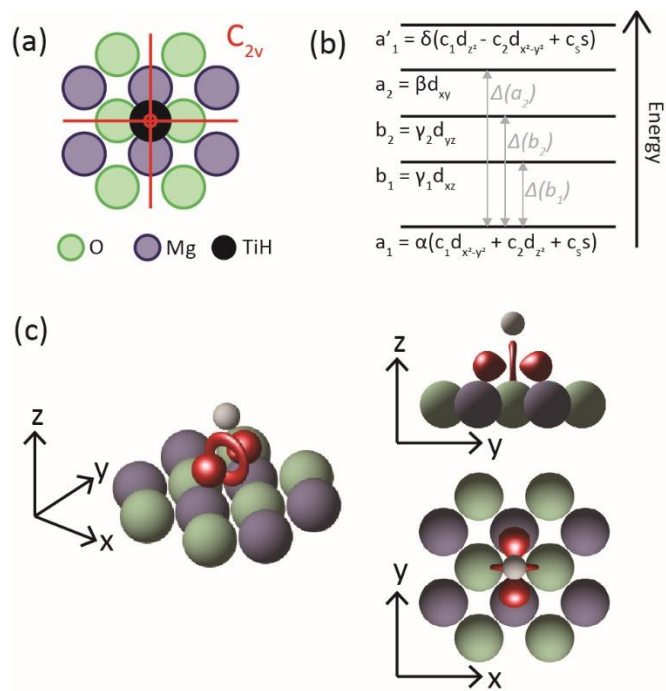

**Figure 4**. Determination of the ground state orbital. (a) Ti is adsorbed on a bridge site with $C_{2v}$ symmetry. (b) Energy diagram for $C_{2v}$ symmetry [16]. The order of the excited states is arbitrary and bears no consequence on the analysis. (c) Iso-surface of the ground-state orbital (red) obtained for $c_s = 0$. Green spheres represent O atoms, blue spheres Mg atoms and the white sphere the H atom on top of Ti.

ASSOCIATED CONTENT

**Supporting Information**. The supplementary information covers details on fitting of ESR spectra, fitting of the hyperfine splitting, anisotropy of the hyperfine splitting in $C_{2v}$ symmetry, point charge model and influence of $c_s$ on the ground state orbital.

**Data and materials availability**. All data presented in this paper are publicly available through Zenodo (18).

AUTHOR INFORMATION

Corresponding Author

*E-Mail: a.f.otte@tudelft.nl

Author Contributions

L.F. and L.M.V. performed the experiments. S.O. supervised the project. All authors analyzed, discussed the results and wrote the manuscript. L.F. and L.M.V. contributed equally to this work.

Notes

The authors declare no competing financial interest.

ACKNOWLEDGMENTS

S.O., L.F. and L.M.V. acknowledge support from the Dutch Research Council (NWO Vici Grant VI.C.182.016) and from the European Research Council (ERC Starting Grant 676895 "SPINCAD"). P.W. acknowledges funding from the Emmy Noether Programme of the DFG (WI5486/1-1).

# Supplementary Material: Experimental Determination of a Single Atom Ground State Orbital through Hyperfine Anisotropy

*Laëtitia Farinacci [†1], Lukas M. Veldman[†1], Philip Willke[2], Sander Otte[1,*]*

1 Department of Quantum Nanoscience, Kavli Institute of Nanoscience, Delft University of Technology, 2628 CJ Delft, the Netherlands.
2 Physikalisches Institut, Karlsruhe Institute of Technology, Karlsruhe, Germany
*E-mail: corresponding author: a.f.otte@tudelft.nl
[†] These authors contributed equally

## Section 1.  Fitting of ESR spectra

### Section 1.1. Fits with modified Fano functions

In order to limit the number of free parameters to fit the hyperfine spectra we recorded for each field direction a reference spectrum on a $^{48}$Ti atom adsorbed on the same bridge site at the same settings as those used for the $^{49}$Ti atom. The reference data is fitted with a modified Fano function [1]:

$$\Delta I = I_0 + I_1 \frac{1+\alpha\delta}{1+\delta^2} \tag{S1}$$

with $\delta = (f - f_0)/(\Gamma/2)$, where $f_0$ is the center frequency of the resonance, $\Gamma$ its width, $I_1$ its intensity, $I_0$ a current offset and $\alpha$ a parameter that accounts for the asymmetry of the resonance.

The data recorded on the hyperfine atom is then fitted with a sum of Fano functions:

$$\Delta I = \tilde{I}_0 + \sum_{i=0}^{i=n-1} \tilde{I}_1 \frac{1+\alpha\delta_i}{1+\delta_i^2} \tag{S2}$$

with $\delta_i = (f - (\tilde{f}_0 + iA))/(\Gamma/2)$, where $\tilde{f}_0$ is the center frequency of the first peak and $A$ the hyperfine splitting parameter. The parameters $\alpha$ and $\Gamma$ are fixed to the values obtained on the reference atom and $n$ is the number of resonances that is determined by the value of the nuclear spin ($n = 6$ for $I = 5/2$ and $n = 8$ for $I = 7/2$). As a result, $\tilde{f}_0$, $A$, $\tilde{I}_0$ and $\tilde{I}_1$ are the free parameters for this fit. We note that the splitting of the peaks are assumed to be identical, which means that we neglect the quadrupole interaction. This assumption is asserted when the peaks are individually resolved, i.e. for relatively large hyperfine splitting. At smaller splitting, the quadrupole interaction could compete with the hyperfine interaction [2] but this remains below our energy resolution.

**Section 1.2. Estimation of the g vector and tip field**

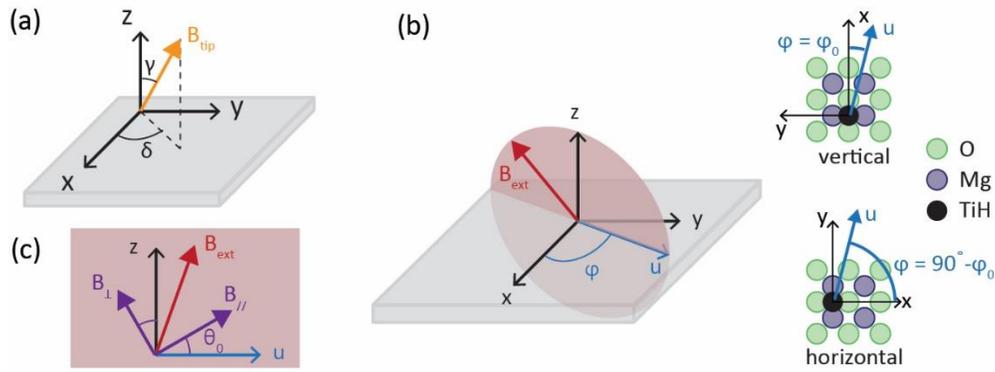

**Figure S1**. Schematics describing the angles considered in this paper. (a) The tip field is described by its amplitude, the angle $\gamma$ it makes with the $z$ axis and its azimuthal angle $\delta$ with respect to the $x$ axis. (b) The measurements span the red plane, defined by the directions of the in-plane and out-of-plane fields. The line at which it intersect the sample plane, with vector director $\vec{u}$ is at an angle $\varphi$ with the $x$ axis. More precisely, for the vertical bridge site $\varphi = \varphi_0$ and for the horizontal bridge site $\varphi = 90° - \varphi_0$. (c) In the rotation plane of the field, to reproduce our experimental observations, we furthermore consider an effective tilt angle $\theta_0$ of the magnet axes with respect to the local crystal field of the atoms.

In this section we evaluate the TiH g factor and the influence of the tip field on its evaluation. We can estimate the strength of the tip field by analyzing the position of the resonances as a function of external field. In the absence of any additional tip field the resonance position for $^{48}$Ti (and the mean position for $^{49}$Ti) is given by

$$f_0 = \frac{\mu_B}{h} \sqrt{(g_\| B_\|)^2 + (g_z B_\perp)^2} \tag{S3}$$

Where $g_\| = g_v$ for the vertical bridge site and $g_\| = g_h$ for the horizontal bridge site. Neglecting any additional field, we obtain for the vertical bridge site: $g_v = 1.688 \pm 0.005$ and $g_z = 2.005 \pm 0.006$ and, for the horizontal bridge site: $g_h = 1.906 \pm 0.009$ and $g_z = 2.017 \pm 0.009$.

Using eq. 3 of the main text these values can be compared to the literature values [1] when taking into account the tilt of our in-plane field with respect to the crystal lattice (see Figure S1) that we estimate to be $\varphi_0 = 14 \pm 2°$. This angle is given by our experimental setup and has been determined in previous experiments [3]. In Table S1 the results of our evaluation are presented, showing first a small variation within our error bars of $\varphi_0$.

| $\varphi_0$ | 12° | 14° | 16° | Kim et al. |
|---|---|---|---|---|
| $g_x$ | 1.698 | 1.702 | 1.706 | 1.653 |
| $g_y$ | 1.897 | 1.894 | 1.890 | 1.917 |
| $g_z$ | 2.011 | 2.011 | 2.011 | 1.989 |

**Table S1**. g-values obtained from our experimental data taking into account an uncertainty of 2° for $\varphi_0$ and compared to literature values [1] (right column).

We find a good agreement with the literature values and the small deviations can be explained by the presence of a small residual tip field, that has been carefully accounted for in [1]. More precisely, using from now on the g values of [1], we consider a field that has a fixed magnetization direction and can switch along its axis so that $\boldsymbol{B}_{tip} \cdot \boldsymbol{B}_{ext} > 0$ [1]. We describe the tip field by its amplitude $|\boldsymbol{B}_{tip}|$ and the angle it makes with respect to the surface normal, $\gamma$, as well as its azimuthal angle with the x axis, $\delta$ (see Figure S1a).

We find for the microtip used for the vertical bridge site data (the error bars correspond to the 95% confidence interval of the fit coefficients)

| $\varphi_0$ | 12° | 14° | 16° |
|---|---|---|---|
| $|\boldsymbol{B}_{tip}|$ | $100 \pm 30$ mT | $110 \pm 40$ mT | $80 \pm 40$ mT |
| $\gamma$ | $101 \pm 4°$ | $111 \pm 4°$ | $106 \pm 7°$ |
| $\delta$ | $91 \pm 3°$ | $89 \pm 1°$ | $90 \pm 2°$ |

**Table S2**. Determination of the tip-field for the micro-tip used to record the data taken on Ti atoms adsorbed on a vertical bridge site; whilst considering our error bars of $\varphi_0$.

And for the horizontal data set we have (we used $\delta = -\varphi_0$ since the value obtained for $g_h$ corresponds to the effective value obtained from Ref. [1] along the field direction)

| $\varphi_0$ | 12° | 14° | 16° |
|---|---|---|---|
| $|\boldsymbol{B}_{tip}|$ | $11 \pm 5$ mT | $12 \pm 5$ mT | $13 \pm 5$ mT |
| $\gamma$ | $25 \pm 155°$ | $15 \pm 105°$ | $5 \pm 60°$ |

**Table S3**. Determination of the tip-field for the micro-tip used to record the data taken on Ti atoms adsorbed on a horizontal bridge site; whilst considering our error bars of $\varphi_0$.

## Section 2.     Fitting of the hyperfine splitting

### Section 2.1. Determination of the hyperfine values

Based on eq. (3) of the main text we fit the data of Fig.3c of the main text with the following function

$$A = \frac{1}{g}\sqrt{\tilde{l}^2 g_{v,h}^2 A_{v,h}^2 + n^2 g_z^2 A_z^2} \tag{S4}$$

Where $\tilde{l}$ and $n$ are the cosine directions of the external field along the $u$ and $z$ axis respectively (see Figure S1b). As mentioned in the main text, we observe a rotation of the data in Fig.3c with respect to the magnet axes. Possible origins for this effective tilt are discussed in the next subsection. We account for it by an effective tilt between the magnet axes and the crystal field axes of the atom. More precisely, we consider an offset angle $\theta_0$ between $(\boldsymbol{B}_\parallel, \boldsymbol{B}_\perp)$ and $(\boldsymbol{u}, \boldsymbol{z})$ (see Figure S1c) and we therefore have

$$\tilde{l} = \cos(\theta - \theta_0) \tag{S5}$$

$$n = \sin(\theta - \theta_0) \tag{S6}$$

where $\tan\theta = \frac{B_\perp}{B_\parallel}$.

The fits in Fig.3c of the main text are based on eq.(S4)-(S6) and show a very good agreement with the experimental data. We obtain for the vertical bridge site $\theta_0 = -6.8° \pm 0.8°$, $A_v = 65.4 \pm 0.7$ MHz, $A_z = 21.7 \pm 1$ MHz and for the horizontal bridge site $\theta_0 = -15° \pm 5°$, $A_h = 23.6 \pm 0.8$ MHz, $A_z = 16.1 \pm 0.8$ MHz.

Taking into account the presence of a tip-field, as determined in the previous section, does not improve the quality of the fits and leads to variations of less than 1 MHz of the fit coefficients. In particular, the presence of a tip field cannot account for the different values of $A_z$. These are most likely due to local variations of the electric field for each atom as also observed in Ref. [4].

To obtain the values of the hyperfine vector along the lattice direction we again have to take into account the tilt of the in-plane field with respect to the crystal lattice. Using eq. (3) of the main text we have

| $\varphi_0$ | 12° | 14° | 16° |
|---|---|---|---|
| $A_x$ | $67 \pm 2$ MHz | $68 \pm 2$ MHz | $69 \pm 2$ MHz |
| $A_y$ | $20.5 \pm 1.5$ MHz | $19 \pm 2$ MHz | $17 \pm 2$ MHz |

**Table S4**. Values for hyperfine splitting along the x and y axis when taking into account the error bars for $\varphi_0$.

As one can see, the uncertainty concerning $\varphi_0$ dominates the error bars for $A_x$ and $A_y$.

### Section 2.2. Possible origins of the rotation

As discussed in the main text and the last section, we observe an offset $\theta_0 \sim 10°$ between the symmetry axis of the measured data and the magnet axes. In this section, we consider several origins for this observation: tip magnetic field, tilt of the STM head, tilt of the sample, local variations in the electric field emerging from inhomogeneities of the substrate, and electrostatic forces emanating from the tip.

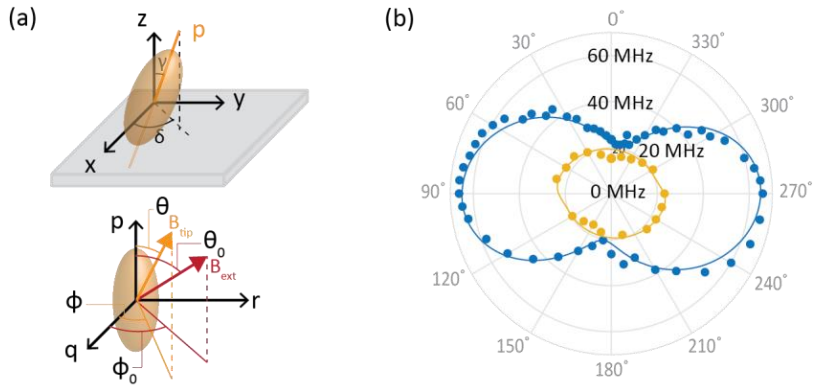

**Figure S2**. (a) Description of the tip field by the Stoner-Wohlfarth model. The orientation of its easy axis $p$ is defined by the angle $\gamma$ that makes with the z axis, and its azimuthal angle $\delta$ with the x axis (top). Around the easy axis, the tip and magnetic fields make an angle $\theta$ and $\theta_0$ with $p$ and have azimuthal angles $\phi$ and $\phi_0$ (bottom). The orientation of $\boldsymbol{B_{tip}}$ is then determined by minimizing the energy of the magnet. (b)

Fitting of the experimental data to account for the rotation with respect to the magnet axes using a tip magnetic field described by the Stoner-Wohlfarth model (note that contrary to Fig.3c of the main text, the values of $A_x$, $A_y$ and $A_z$ are not fit parameters but fixed).

In order to investigate if the tip magnetic field could be at the origin of the observed rotation between the data of Fig.3c and the magnet axes we modelled the tip field with a Stoner-Wohlfarth model. This model allows for the tip magnetization to deviate from its main axis and therefore covers a wider range of magnet behaviors in an external field than the model considered in the previous section. More precisely, the orientation of the easy axis of the magnet, $p$, with respect to the (x,y,z) basis is characterized by its angle $\gamma$ with the z axis, and its azimuthal angle $\delta$ with respect to the $x$ axis (see Fig.S2a). The behavior of the magnet is assumed to be isotropic around this axis. Considering a basis (p,q,r) around this axis the orientation of the magnet and external fields are characterized by their angles with respect to the $p$ axis, $\theta$ and $\theta_0$, and their azimuthal angles, $\phi$ and $\phi_0$, respectively. The energy E of the system is then given by

$$E = \frac{V\,B_{tip}}{2}[-H_k \cos^2\theta - 2B_{ext}(\sin\theta_0 \sin\theta \cos\phi_0 \cos\phi$$
$$+ \sin\theta_0 \sin\theta \sin\phi_0 \sin\phi + \cos\theta_0 \cos\theta)] \tag{S7}$$

where V is the volume of the magnet and $H_k$ is the anisotropy field. The first term is the magnetic anisotropy and the second the Zeeman energy. The orientation of the tip field is then obtained by determining the minima of E, i.e. by solving $\frac{\partial E}{\partial \phi} = 0$ and $\frac{\partial E}{\partial \theta} = 0$. When two minima are found we weight each solution by their Boltzmann population at 1K.

Using this model, we can estimate which tip field would be needed to reproduce the rotation observed between the experimental data and the magnet axes. More precisely, we fit the experimental data using eq.(2) and (3) of the main text, where we fix $(g_x, g_y, g_z) = (1.653, 1.917, 1.989)$, and $(A_x, A_y, A_z) = (68, 18, 19)$. The total magnetic field is the sum of the external field (with the out-of-plane and in-plane components aligned with the z and u axes) and tip field - the latter being determined via the Stoner-Wohlfarth model. We estimate that the volume of the magnet is that of a sphere with a radius of 2 Å and set $H_k = 100$ T (we find empirically that this value of $H_k$ gives better fits to our experimental data). The fits for both adsorption sites are shown in Fig.S2b and yield, for the vertical bridge site, $B_{tip} = 0.565 \pm 0.125$ T, $\gamma = 82 \pm 3°$ and $\delta = -67 \pm 2°$; and for the horizontal bridge site, $B_{tip} = 0.270 \pm 0.670$ T, $\gamma = 13 \pm 22°$ and $\delta = -6 \pm 106°$. The fits demonstrate that a tip field can induce an apparent rotation of the data with respect to the magnet axes but do not completely reproduce the shape of the experimental data. Additionally, the tip fields corresponding to the fits would shift the ESR resonance frequencies $f_0$ substantially and therefore are too high to be realistic — by a factor ~ 5 for the vertical bridge site and ~ 20 for the horizontal bridge site (see section S1.2).

We identify a number of other possible origins for the tilt in the measurements for Fig 3c. We estimate the uncertainty of the STM head tilt with respect to the external field to be <5° based on the geometry of the STM design. The misalignment of the sample with respect to the scanning piezo was measured to be ~-0.2° in each direction. We therefore expect that although these errors could accumulate, macroscopic origins alone should be insufficient to explain the magnitude of the observed rotation.

We also consider possible origins of a microscopic nature. Local variations in the electric field emerging from inhomogeneities of the substrate have been linked to variations in the g-factor of TiH atoms adsorbed on O-sites of the MgO lattice. The g-factor was found to be especially susceptible to changes in the in-plane direction of the electrostatic field leading to variations up to 15% [4]. This indicates that local charges play a large role in the crystalline environment the atom experiences.

Next to that, due to the anisotropic shape of the tip, the in-plane components of van der Waals and electrostatic forces emanating from the tip could move the atom slightly with respect to the crystal lattice. Similar in-plane tilting effects have been observed for molecules adsorbed on AFM tips [5] and in mechanical bond breaking experiments [6], and has been used for lateral atomic manipulation [7,8]. The driving mechanism behind ESR-STM itself has been attributed to the electric field inducing movement of the adatom via a piezo crystalline effect in the MgO [9,10].

While we cannot definitively identify the origin of the observed rotation, we believe that the effects discussed above may each contribute to its explanation. Independent of the origin, the hyperfine values and their error bars determined by the minima and maxima of the hyperfine splitting in Figure 3c remain unaffected.

## Section 3.    Anisotropy of the hyperfine splitting in $C_{2v}$ symmetry

The anisotropy of the g and A vectors in $C_{2v}$ symmetry is given by eq. (7) of the main text where the functions $f_x$, $f_y$ and $f_z$ are defined as follows [11 p.382]

$$f_x = -\frac{4}{3}(c_1 + \sqrt{3}c_2)^2 K_2 + \frac{2}{3}\left[(c_1 - \sqrt{3}c_2)^2 K_3 + 4c_1^2 K_1\right] + \frac{2}{7}(c_1^2 - c_2^2 - 2\sqrt{3}c_1 c_2) - \frac{4\sqrt{3}}{7} c_1 c_2 K_1$$

$$+ \frac{\sqrt{3}}{7}(\sqrt{3}c_1 + c_2)(c_1 - \sqrt{3}c_2) K_3 \tag{S8}$$

$$f_y = -\frac{4}{3}(c_1 - \sqrt{3}c_2)^2 K_3 + \frac{2}{3}\left[(c_1 + \sqrt{3}c_2)^2 K_2 + 4c_1^2 K_1\right] + \frac{2}{7}(c_1^2 - c_2^2 + 2\sqrt{3}c_1 c_2) + \frac{4\sqrt{3}}{7} c_1 c_2 K_1$$

$$+ \frac{\sqrt{3}}{7}(\sqrt{3}c_1 - c_2)(c_1 + \sqrt{3}c_2) K_2 \tag{S9}$$

$$f_z = -\frac{16}{3} c_1^2 K_1 + \frac{2}{3}\left[(c_1 + \sqrt{3}\, c_2)^2 K_2 + (c_1 - \sqrt{3}\, c_2)^2 K_3\right] - \frac{4}{7}(c_1^2 - c_2^2)$$

$$- \frac{\sqrt{3}}{7}(\sqrt{3}\, c_1 + c_2)(c_1 - \sqrt{3}\, c_2) K_3 - \frac{\sqrt{3}}{7}(\sqrt{3}\, c_1 - c_2)(c_1 + \sqrt{3}\, c_2) K_2 \tag{S10}$$

We calculate the values of $\Delta A_x$, $\Delta A_y$ and $\Delta A_z$ for parameter sets $(P, \alpha, c_1, c_2, c_s)$ where P spans $[0; -200]$ MHz, $\alpha$ spans $[0;1]$, $c_s$ spans $[0; 0.8]$ for which we only calculate sets in increments of 0.2. $c_1$ and $c_2$ are calculated via the normalization equation $c_1^2 + c_2^2 + c_s^2 = 1$. An angle $\chi$ is defined as $\tan \chi = \frac{c_2}{c_1}$, where $\chi$ spans $[0°; 360°]$. The calculation is performed in the following way: first the values of $K_1$, $K_2$, and $K_3$ are calculated via eq. (4)-(6) of the main text and then $\Delta A_x$, $\Delta A_y$ and $\Delta A_z$ are obtained from eq. (7), (S8)-(S10).

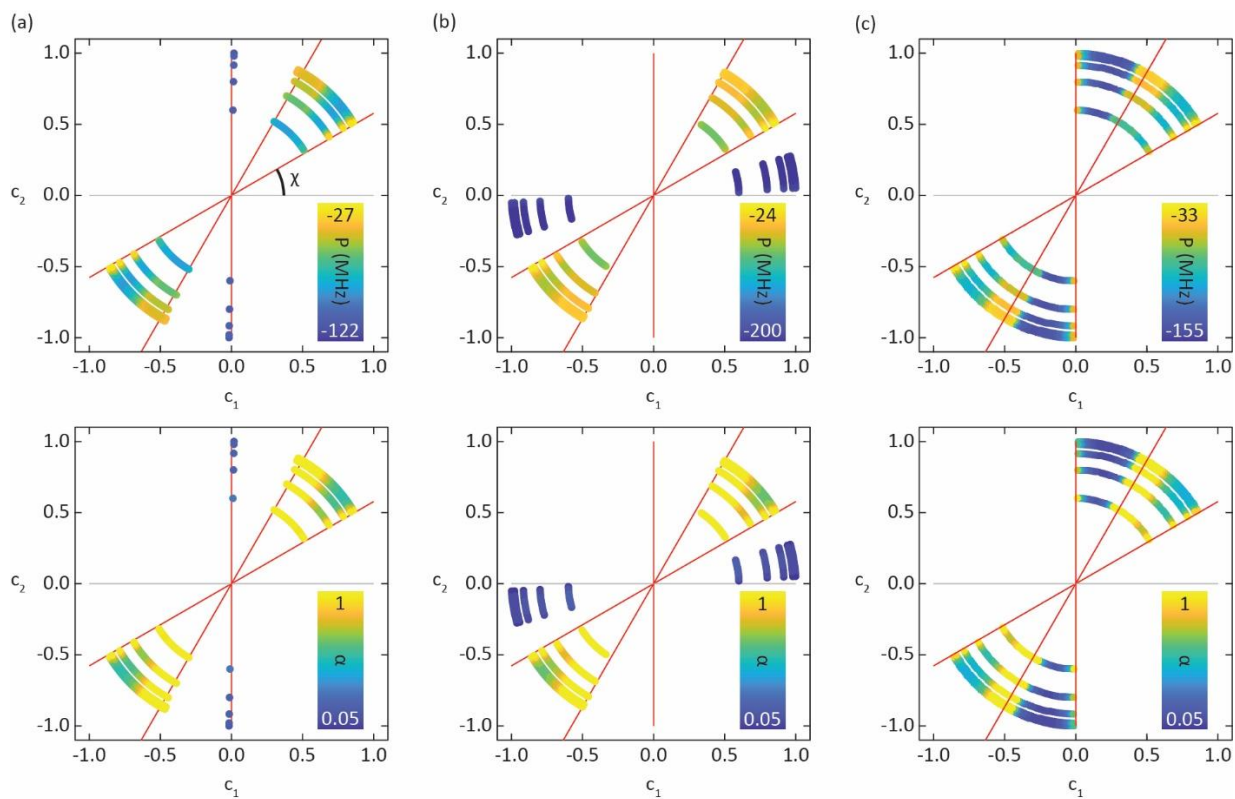

**Figure S3**. Sets of parameters $(P, \alpha, c_1, c_2, c_s)$ that can give rise the observed anisotropy of the hyperfine splitting. (a-c) considers additionally variations of **g** (a) Values as reported in Ref. [1]: $(g_x, g_y, g_z) = (1.653, 1.917, 1.989)$. (b) $(g_x, g_y, g_z) = (1.655, 1.898, 2.013)$. (c) $(g_x, g_y, g_z) = (1.651, 1.936, 1.965)$.

A parameter set is considered to be a valid solution if the values obtained for $\Delta A_x$, $\Delta A_y$ and $\Delta A_z$ are within $\pm 4$ MHz of the experimental values. We plot in Fig. S3a the data points that correspond to such valid solutions: each solution is represented with a point whose coordinates are $(c_1, c_2)$ and the value of $c_s$ determines the distance between this point and the origin (because of the normalization equation) - as a result the different concentric cycles correspond to the different values of the $c_s$ parameter. The two remaining parameters, $P$ and $\alpha$, are represented by the color of the points in upper and lower graphs, respectively. We ensured the stability of our model against the uncertainty of **g** reported in literature and Fig. S3b-c, shows two examples of the results of the calculations performed when considering a combination of extrema/minima of the components of **g** (we performed the calculation for all 8 possible combinations).

As one can see, the value of $c_s$ does not discriminate between different $\chi$ values but rather renormalizes the values of $\alpha$ and $P$. The two lines corresponding to $\chi = 30°$ and $\chi = 60°$ indicated by red lines are robust against variations of **g** and correspond to reasonable values for $\alpha$ and $P$. Indeed, $\alpha$ quantifies the hybridization of the d levels with ligands orbitals and, since, we assume this effect to be minor, $\alpha$ should be close to 1. On the other hand, $P$ scales with $\langle r^{-3} \rangle$ and large values of $P$ correspond to orbitals with a very small spatial extent, from literature values [11 – Table 9.13 p.359] we expect $P \sim -78$ MHz for a $Ti^{3+}$.

## Section 4. Point charge model

| Ion | Δ electrons (no.) | X(Å) | Y(Å) | Z (Å) |
|---|---|---|---|---|
| H | 1 | 0 | 0 | 1.8 |
| O | 2 | -1.45 | 0 | -1.6 |
| O | 2 | 1.45 | 0 | -1.6 |
| Mg | -2 | 0 | -1.45 | -1.6 |
| Mg | -2 | 0 | 1.45 | -1.6 |

**Table S5**. Point charge model used to identify the ground state orbital. Shown are the local charges as well as their position in the $(x, y, z)$ coordinate system centered around the Ti atom (see Figure 1 of the main text).

To discriminate between the different solutions shown in Fig. S3, we use additionally a point charge model defined from Table S5. The point charge model allows to distinguish solutions that yield the correct hyperfine values (Fig. S3), but are unlikely ground states, since the orbital charges are pointing in unfavorable directions of the surrounding crystal field. Each charge $q_i$ at a position $(x_i, y_i, z_i)$ yields a potential

$$V_i = \frac{q_i}{\sqrt{(x-x_i)^2+(y-y_i)^2+(z-z_i)^2}} \tag{S11}$$

So that the total Coulomb energy for an electron in an orbital $|\psi\rangle$ is $E_C = -e\langle\psi|\sum_i V_i|\psi\rangle$.

For each set of parameters that yields correct values for the anisotropy of the hyperfine vector we calculate the corresponding ground state orbital

$$|\psi\rangle = c_1 d_{x^2-y^2} + c_2 d_{z^2} + c_s 4s \tag{S12}$$

where $d_{x^2-y^2}$, $d_{z^2}$ and 4s are the spherical harmonics for which the radial parts verify

$$R_{3,2} = \frac{4}{81\sqrt{6}} \left(\frac{Z_{3d}}{a_0}\right)^{3/2} \rho^2 \exp\left(-\frac{\rho}{3}\right) \tag{S13}$$

$$R_{4,0} = \frac{1}{96} Z_{4s}^{3/2} \left[24 - \frac{26\rho}{2} + 12\left(\frac{\rho}{2}\right)^2 - \left(\frac{\rho}{2}\right)^3\right] \exp\left(-\frac{\rho}{4}\right) \tag{S14}$$

where $\rho = Zr/a_0$, with $a_0$ being the Bohr radius and $Z_{3d}$ ($Z_{4s}$) the effective nuclear charge for the $3d$ ($4s$) shell.

And we have for the angular parts

$$Y_{x^2-y^2} = \frac{\sqrt{15}}{4\sqrt{\pi}} \frac{x^2-y^2}{r^2} \tag{S15}$$

$$Y_{z^2} = \frac{\sqrt{5}}{4\sqrt{\pi}} \frac{3z^2-r^2}{r^2} \tag{S16}$$

$$Y_{4s} = \frac{1}{\sqrt{4\pi}} \tag{S17}$$

Furthermore, we have (see main text)

$$P = g_0 \, g_N \, \mu_N \, \mu_B \, \langle r^{-3} \rangle \tag{S18}$$

Neglecting for simplicity any contribution of $c_s$, the radial extent of the orbital can be calculated from the radial wave-function for $d$ orbitals [see eq. S(13)]

$$\langle r^{-3} \rangle = \int_0^\infty [R_{3,2}(r)]^2 r^{-3} r^2 dr \tag{S19}$$

And we obtain

$$\langle r^{-3} \rangle = \frac{Z^3}{81 a_0^3} \tag{S20}$$

Therefore, for each set of solutions determined after the previous step, we calculate the value of $Z_{3d}$ using the value of the $P$ parameter and eq. (S18) and (S20). The effective nuclear charge for the $4s$ orbital is then adjusted so that the ratio $\frac{Z_{3d}}{Z_{4s}}$ equals the one given in literature [12].

The calculation is performed using a grid in the $(x, y, z)$ space that spans $[-4a_0 : 4a_0]$ in each direction and with a spacing of $0.1 a_0$ between points. We ensure robustness of the results by varying the position of the Ti atom with respect to the crystal lattice along the z direction in the range of 20%. The position of the Mg and O atoms are determined experimentally by atomic resolution images and the one of the H atom is set according to [13].

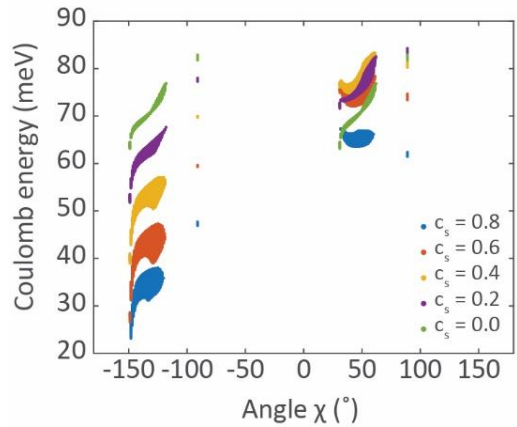

**Figure S4**. Coulomb energy for the different sets of parameters that correctly describe the hyperfine anisotropy.

In Figure S4 we show the Coulomb energy calculated for the sets of parameters shown in Figure S4a. Each color represents a different value of $c_s$ and the multiplicity of points for given $\chi$ and $c_s$ values corresponds to the multiple sets of candidates that contains these values. As one can see, decreasing $c_s$ leads to a systematic decrease in the Coulomb energy. While this can be easily explained by the smaller radial extent of the $4s$ orbital with respect to the $3d$ orbitals, the calculation suggests that the minimal solution corresponds to an electron only localized in the $4s$ orbital which is unrealistic. The point charge model therefore does not allow to determine with certainty the value of $c_s$. However, it allows us to clearly identify the ground state orbital for each value of $c_s$. As we consistently find a minimum at $\chi = -150°$ for every value of $c_s$ we conclude that the relative mixture between the *d* orbitals is not affected by the addition of $c_s$.

## Section 5. Influence of $c_s$ on the ground state orbital

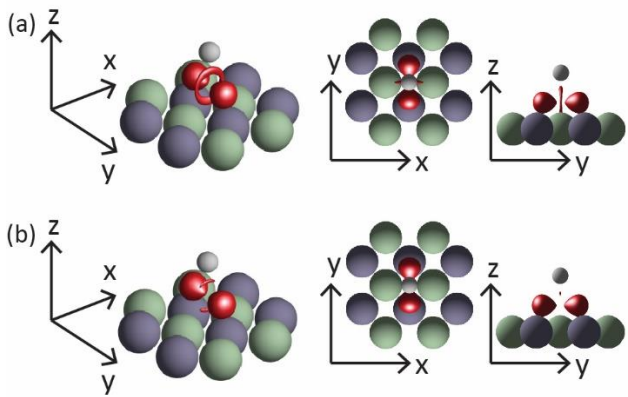

**Figure S5**. Influence of $c_s$ on the ground state orbital. Solution obtained for $c_s = 0.2$ (a) and $c_s = 0.4$ (b).

In the main text, we show the solution obtained for $c_s = 0$. In Figure S5, we show the optimal solutions for $c_s = 0.2$ (a), which is most likely an upper boundary for the admixture of the $4s$ orbital, as well as for $c_s = 0.4$ (b), which is unrealistic but help us better capture the influence of the parameter. The admixture of the $4s$ orbital influences only marginally the shape of the orbital: it mostly reduces the size of the central ring that points towards the neighboring O atoms. We ensured that these results are robust against variations of the g vector within the error bars given in Ref. [1].